\documentclass[letterpaper]{mc2023}
\usepackage{tabls}
\usepackage{cites}
\usepackage{epsf}
\usepackage{appendix}
\usepackage{ragged2e}
\usepackage[top=1in, bottom=1in, left=1in, right=1in]{geometry}
\usepackage{enumitem}
\setlist[itemize]{leftmargin=*}
\usepackage{caption}
\usepackage{algorithm}
\usepackage{algorithmic}
\usepackage{marvosym}
\usepackage{wrapfig}
\newcommand\mcdc{MC\Lightning DC}

\title{iQMC: Iterative Quasi-Monte Carlo \\
for k-Eigenvalue Neutron Transport Simulations}
\author{%
  \textbf{S.~Pasmann$^1$, I.~Variansyah$^2$, C.T.~Kelley$^3$, and R.G.~McClarren$^1$}\vspace{3pt} \\
  $^1$University of Notre Dame  \\
  Department of Aerospace and Mechanical Engineering\\ 
  Fitzpatrick Hall, Notre Dame, IN 46556 \vspace{6pt}\\ 
  $^2$Oregon State University \\
  Department of Nuclear Science and Engineering\\
  1791 SW Campus Way, Corvallis, OR \vspace{6pt}\\
  $^3$North Carolina State University \\ 
     Department of Mathematics\\ 3234 SAS Hall, Box 8205\\ Raleigh NC 27695-8205\vspace{6pt} \\ 
  \url{spasmann@nd.edu}, \url{variansi@oregonstate.edu}, \url{ctk@ncsu.edu}, \url{rmcclarr@nd.edu}
}
\newcommand{\authorHead}{Authors' names, use et al.~if more than three}
\newcommand{\shortTitle}{Short One Line Paper Title}
\begin{document}
\maketitle
\justify 
\parskip 6pt plus 1 pt minus 1 pt

\begin{abstract}
The Iterative Quasi-Monte Carlo method, or iQMC, replaces standard quadrature techniques used in deterministic linear solvers with Quasi-Monte Carlo simulation for more accurate and efficient solutions to the neutron transport equation. This work explores employing iQMC in the Monte-Carlo Dynamic Code (\mcdc) to solve k-eigenvalue problems for neutron transport with both the standard power iteration and the generalized Davidson method, a Krylov Subspace method. Results are verified with the 3-D, 2-group, Takeda-1 Benchmark problem.
\end{abstract}
\vspace{6pt}

\section{INTRODUCTION} 
\label{sec:intro}

\begin{figure}[!htb]
  \centering
  \includegraphics[scale=0.66]{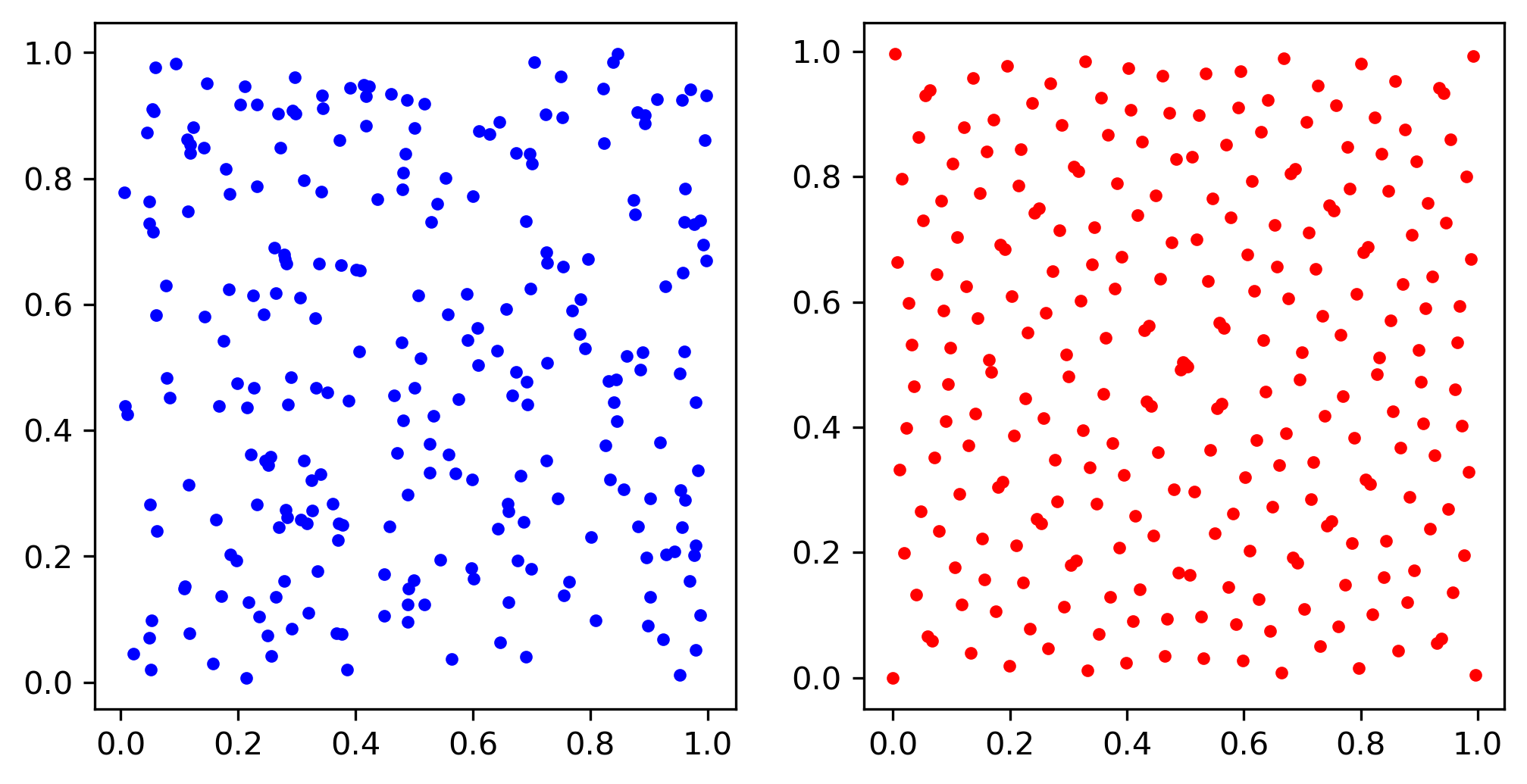}
  \caption{256 points generated in the unite square from a pseudo-random number generator (left) and the Sobol Sequence (right).}
  \label{fig:lds}
\end{figure}

Quasi-Monte Carlo (QMC) is the use of low-discrepancy sequences (LDS) in place of pseudo-random number generators in Monte Carlo (MC) simulation. Figure~\ref{fig:lds} shows how LDS use quasi-random or deterministic algorithms to sample the phase-space in a more efficient manner than random samples. This leads to a theoretical convergence rate of $O(N^{-1})$ compared to the standard MC convergence rate of $O(N^{-1/2})$, where $N$ is the number of samples \cite{bickel2009monte}. Despite this benefit, QMC has largely been ignored by the particle transport community. This is primarily because the quasi-random or deterministic nature of the LDS breaks the Markovian assumption needed to model the particle random walk process \cite{spanier1995quasi}. 

Our method circumvents the need to model particle scattering by treating the scattering and fission terms as internal fixed sources, reducing the (Quasi) Monte Carlo simulation to a particle ray-trace. The scattering and fission sources are then iterated upon in an outer loop using deterministic iterative methods. In previous work, we applied this iterative Quasi-Monte Carlo method (iQMC) to several 1D fixed source problems and achieved the expected $O(N^{-1})$ convergence rate across all test problems \cite{Pasmann2023}. The iterative process was also accelerated with the use of Krylov Linear Solvers in place of the standard source iteration. The Krylov solvers converged with as few as $1/8$ the number of iterations required by the source iteration.

We investigate the application of iQMC to k-eigenvalue or criticality problems for neutron transport. In addition to the standard power iteration, we employ the generalized Davidson method, a Krylov subspace method which has been shown to outperform power iteration in similar k-eigenvalue problems~\cite{Subramanian2011}. iQMC verification results come from the Takeda-1 Benchmark problem \cite{Takeda1991}, a 3-D, 2-group, reactor implemented in the Monte Carlo Dynamic Code (\mcdc) \cite{variansyah_mc23_mcdc}.

\section{METHODOLOGY} 
\label{sec:methods}

\subsection{Linear Operators for use with Iterative Methods for k-Eigenvalue Problems}
We begin with the time-independent form of the Boltzmann Neutron Transport Equation~(\ref{eq:nte_eigen}) for angular flux $\psi(\mathbf{r}, \mathbf{\Omega}, E)$ with isotropic scattering, where $\mathbf{r}$ and $\mathbf{\Omega}$ are the spatial and angular vectors respectively and $E$ is the neutron energy \cite{Lewis1984}:
\begin{multline}
    \label{eq:nte_eigen}
    \Omega\cdot\nabla\psi(\mathbf{r},\mathbf{\Omega},E)+\Sigma_{t}(\mathbf{r},E)\psi(r,\mathbf{\Omega},E) = \\ \int_{0}^{\infty}dE^{\prime}\int_{4\pi}d\mathbf{\Omega}^{\prime}\Sigma_{s}(\mathbf{r},\mathbf{\Omega}^{\prime}\rightarrow\mathbf{\Omega},E^{\prime}\rightarrow E)\psi(\mathbf{r},\Omega^{\prime},E^{\prime}) \\ 
    + \frac{\chi(\mathbf{r},E)}{4\pi k_\mathrm{eff}}\int_{0}^{\infty}dE^{\prime}\nu\Sigma_{f}(\mathbf{r},E^{\prime})\int_{4\pi}d\mathbf{\Omega}^{\prime}\psi(\mathbf{r},\mathbf{\Omega}^{\prime},E^{\prime}).
\end{multline}

Here, $\chi(\mathbf{r}, E)$ is the fission energy distribution, $\Sigma_t(\mathbf{r}, E)$ the total macroscopic cross section, $\Sigma_f(\mathbf{r}, E^{\prime})$ the fission cross section,  $\Sigma_{s}(\mathbf{r},\Omega^{\prime}\rightarrow\Omega,E^{\prime}\rightarrow E)$ the scattering cross section, $\nu$ is the number of neutrons produced per fission event, and $k_\mathrm{eff}$ the effective neutron multiplication factor. Defining the scalar flux $\phi(\mathbf{r},E)$ as the integral over angle of the angular flux,  $\phi(\mathbf{r},E)=\int_{4\pi}d\mathbf{\Omega}^{\prime}\psi(\mathbf{r},\mathbf{\Omega}^{\prime},E)$.
Equation~(\ref{eq:nte_eigen}) can be represented in linear operator form as \cite{hamilton2011numerical}:
\begin{equation}
\label{eq:lin_op}
L\phi = S\phi + \frac{1}{k_\mathrm{eff}}F\phi,
\end{equation}
where $L$ is the streaming operator, $S$ is the scattering operator, and $F$ is the fission operator. To combine iterative methods with QMC we define linear operators from QMC simulation that can be used in place of matrices in the iterative method. 

\subsection{Power Iteration}
\label{sec:PI}
The fixed-point power iteration is a common iterative method for k-eigenvalue problems in neutron transport \cite{warsa2004krylov}. Mathematically, the power iteration searches for solutions to the standard eigenvalue problem,
\begin{equation}
\label{eq:std_eig}
A\phi = k_\mathrm{eff}\phi.
\end{equation}
Operator $A$ in Equation~(\ref{eq:std_eig}) can be defined in terms of the streaming, scattering, and fission operators from Equation~(\ref{eq:lin_op}) as:
\begin{equation}
\label{eq:A}
A = (I-L^{-1}S)^{-1}L^{-1}F.
\end{equation}
Unlike traditional numerical solvers, we do not need to explicitly form the matrix $A$. Instead we define functions that use the Quasi-Monte Carlo simulation (discussed in Section~\ref{sec:qmc_sweep}) to compute the actions of the matrix on an input scalar flux $\phi$. These ``matrix-vector product functions'' act as linear operators to be called with each iteration. Note the structure of $A$ requires a complete source iteration per power iteration. The use of iQMC to solve source iteration problems is described in \cite{Pasmann2023}.

\subsection{The Davidson Method}
\label{sec:davidsons}
 The power iteration is a fixed-point iterative method which generates the next iterate based solely on the information from previous iterate. This typically requires a precise solve of the scattering source to converge. Moreover, the power iteration has been shown to struggle in problems where the two largest eigenvalues are close to one another, also known as a high dominance ratio. We also consider the generalized Davidson method (referred to from now on as the Davidson method), which was first developed in 1975 to compute the $l$ extreme (largest or smallest) eigenpairs ($\lambda, y$) and whose convergence is not limited by high dominance ratios \cite{Davidson1975,warsa2004krylov}. Unlike the power iteration, the Davidson method stores a resultant vector from each iteration and uses the collective information to generate the next vector. The Davidson method typically requires far fewer iterations to converge than the power iteration. Mathematically, the Davidson method searches for solutions to the generalized eigenvalue problem of the form
\begin{equation}
\label{eq:gen_eig}
B\phi = \frac{1}{k_\mathrm{eff}}C\phi.
\end{equation}
We can represent the linear operators $B$ and $C$ in Equation~(\ref{eq:gen_eig}) in terms of the streaming, scattering, and fission operators
\begin{equation}
\label{eq:B}
B = L^{-1}F, \qquad \text{and}\qquad  
C = (I-L^{-1}S).
\end{equation}
An important consideration for the Davidson method is the choice of an appropriate preconditioner $M$. $M$ is an approximation to $C$ and in iQMC this takes the form of a specified number of QMC transport sweeps of the scattering source. The orthogonalization routine called on the Krylov subspace is another important consideration. iQMC uses Modified Gram Schmidt as it has been shown to best uphold orthogonalization, particularly after a large number of iterations \cite{kelley1995iterative}. Finally, as the number of iterations grows, and consequently the Kyrlov subspace, memory constrains can become of primary concern. To combat this issue, we introduce a restart parameter $m$, that represents the maximum allowable size of the subspace. If the subspace exceeds $m$ vectors, the procedure reduces the subspace to latest iterate and begins again. An outline of the Davidson method for the generalized eigenvalue problem taken from \cite{Subramanian2011}, adjusted to compute only the largest eigenpair, can be seen in Algorithm~\ref{alg:davidson}.\\
\begin{minipage}{0.66\textwidth}
   \begin{algorithm}[H]
    \caption{Generalized Davidson Method} 
    \label{alg:davidson}
    \begin{algorithmic}
            \STATE Def Davidson($\phi_0$, $B$, $C$, $m$)
                  \STATE $i=0$
                  \STATE $V_0 = \phi_0 / \lVert \phi_0 \rVert_H$
           \WHILE{Not Converged \do}
                        \STATE $B_i = V_{i}^{H}BV_{i}$
                        \STATE $C_i = V_{i}^{H}CV_{i}$
                        \STATE Solve $B_{i}y=\lambda_{i}C_{i}y_{i}$
                        \STATE $\phi_{i} = V_{i}y_{i}$
                        \STATE $r_i = B\phi_{i} - \lambda_{i}C{i}\phi_{i}$
                        \IF{Convergence}
                            \STATE Exit
                        \ENDIF
                        \STATE $t_{i} = Mr_{i}$
                        \IF{$\mathrm{dim}(V_{i}) \le m-1$}
                            \STATE $V_{i+1} = \mathrm{MGS}(V_{i},t_{i})$
                        \ELSE
                            \STATE $V_{i+1} = \mathrm{MGS}(u_{i},t_{i})$
                        \ENDIF
                        \STATE $i = i+1$
                    \ENDWHILE
    \end{algorithmic} 
   \end{algorithm}
\end{minipage}

\subsection{Fixed-Seed Quasi Monte Carlo Sweep}
\label{sec:qmc_sweep}
Quasi-Monte Carlo simulation forms the basis of the linear operators $A$, $B$, and $C$. Whereby removing the need to explicitly model the scattering and fission processes, the system can be treated as a pure-absorber where each particle is given an angle $\mathbf{\Omega}$, position $\mathbf{r}$, and statistical weight $w$. The initial position and angle are generated from the LDS, while the initial weight is calculated via the RHS of Equation~(\ref{eq:nte_eigen}).

After a particle is born, it is traced or \textit{swept} straight out of the volume. The sweep is greatly enhanced with the use of a continuous weight reduction technique which, reduces the statistical weight of the particle per path length traveled. Then a path-length tally estimator is used to compute the spatially-averaged scalar flux in each zone of the defined mesh. iQMC utilizes a multigroup energy discretization, consequently each simulated particle can now represent all energy groups. As the particle is traced out of the volume, each energy group can be attenuated in parallel \cite{Pasmann2023}. 

Figure~\ref{fig:lds} shows the low-discrepancy nature of QMC which ensures a well-distributed sampling of the phase-space even with a relatively few number of samples compared to pseudo-random sampling. This allows the use of fixed-seed approach in the iterative method, every time the QMC Sweep function is called the LDS is reset to the beginning of the sequence. This provides iQMC with  good convergence characteristics so long as the number of particle histories per iteration is greater than the number of spatial cells in the mesh.


\section{IMPLEMENTATION} 
\label{sec:implement}

iQMC was implemented and verified in the Monte Carlo Dynamic Code (\mcdc), a performant, scalable, and machine-portable Python-based Monte Carlo neutron transport software currently in development in the Center for Exascale Monte-Carlo for Neutron Transport (CEMeNT) \cite{variansyah_mc23_mcdc}. \mcdc{} takes advantage of \textit{Numba} \cite{lam2015numba}, a just-in-time compiler for scientific computing in Python which has been shown to provide 56-212x speedups over pure Python on various reactor problems \cite{variansyah_mc23_mcdc}. All experiments were run on 64 cores from Intel Xeon E5-2680 processors, which run at a clockspeed of 2.5GHz, at the University of Notre Dame Center for Research Computing.

\section{RESULTS} 
\label{sec:results}

\begin{figure}[!htb]
  \centering
  \includegraphics[scale=0.66]{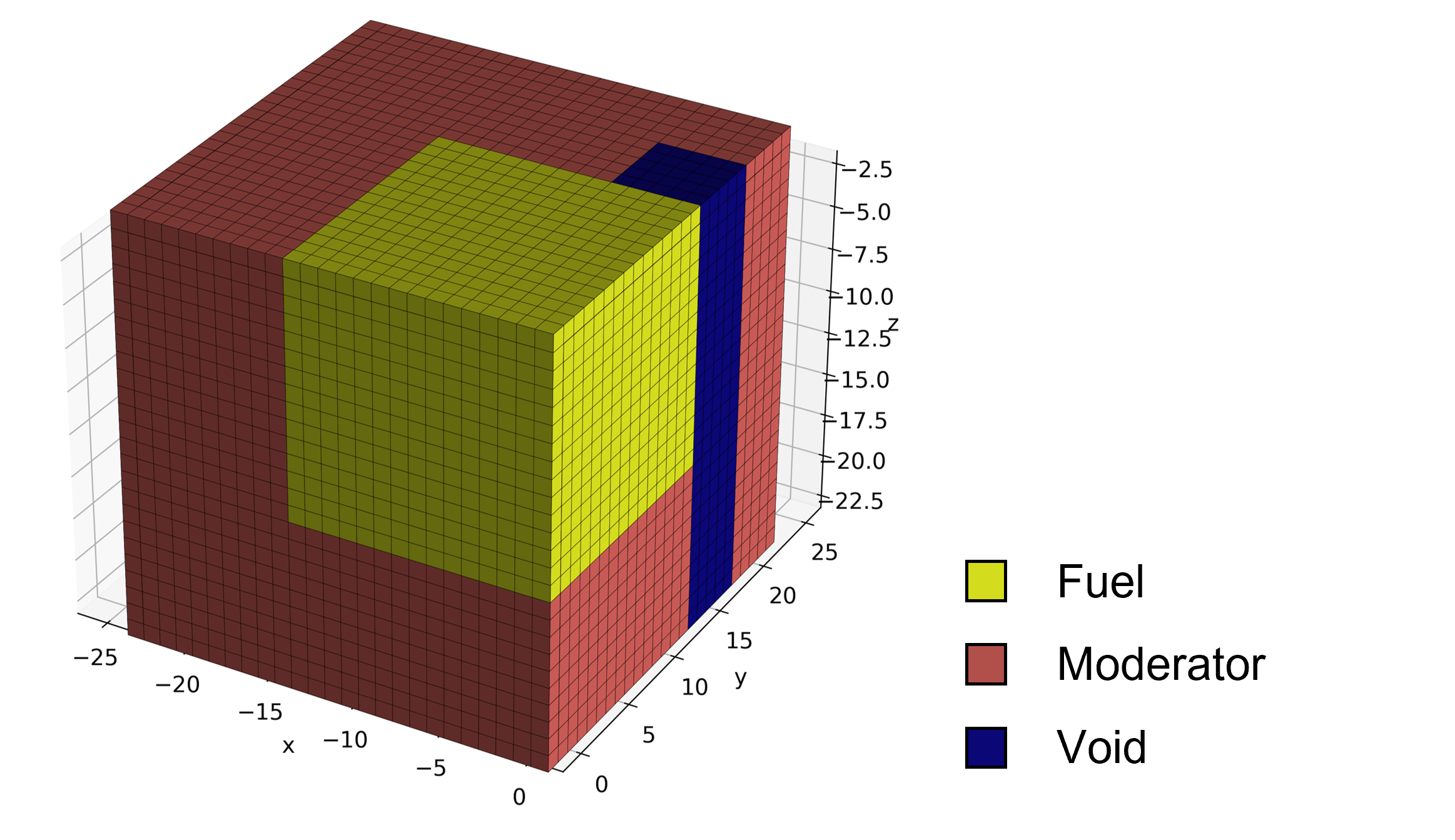}
  \caption{The Takeda-1 Benchmark with removed (void) control rod and 25x25x25 iQMC uniform Cartesian mesh.}
  \label{fig:takeda1_material}
\end{figure}

\begin{figure}[!htb]
  \centering
  \includegraphics[scale=0.75]{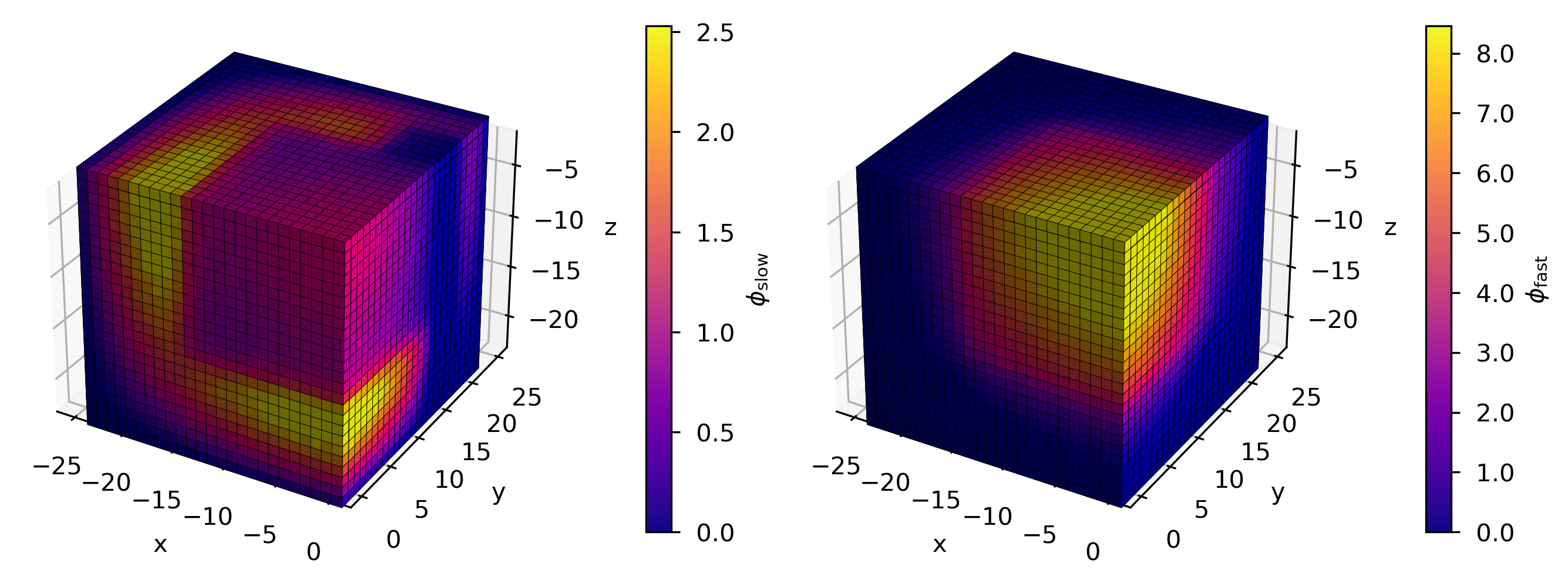}
  \caption{iQMC scalar flux results for slow (left) and fast (right) energy groups.}
  \label{fig:takeda1_flux}
\end{figure}

The Takeda-1 Benchmark problem is a 3-D, 2-group, quarter-core light-water reactor problem. Figure~\ref{fig:takeda1_material} shows a depiction of the reactor geometry. The problem has two variations, the first contains an inserted control rod and the second the control rod is removed (void), cross section data for both cases can be viewed in \cite{Takeda1991}. The reference solution for both $k_\mathrm{eff}$ and point-wise scalar flux $\phi$ come from a high fidelity Monte-Carlo simulation in \mcdc. Our Monte-Carlo simulation obtained a $k_\mathrm{eff} = 0.97779 \pm 3\times10^{-5}$ which agrees well with the reference $k_\mathrm{eff} = 0.9778 \pm 5\times10^{-4}$ (the Monte-Carlo solution) from Takeda et al. \cite{Takeda1991}.

 iQMC results utilize a 25x25x25 uniform Cartesian mesh which has been shown to be adequate for other methods \cite{Criekingen2007}. Both the power iteration and Davidson method defined convergence as $\Delta k_\mathrm{eff}/k_\mathrm{eff} \le 1\times10^{-4}$. Additionally, the internal source iteration of the power iteration converged to $\Delta \phi/\phi \le 1\times10^{-4}$ while the Davidson method was run with 3 preconditioner sweeps. Figure~\ref{fig:takeda1_flux} shows scalar flux results from iQMC with the power iteration, separated into slow and fast groups. 
 
\begin{figure}[!htb]
  \centering
  \includegraphics[width=\textwidth]{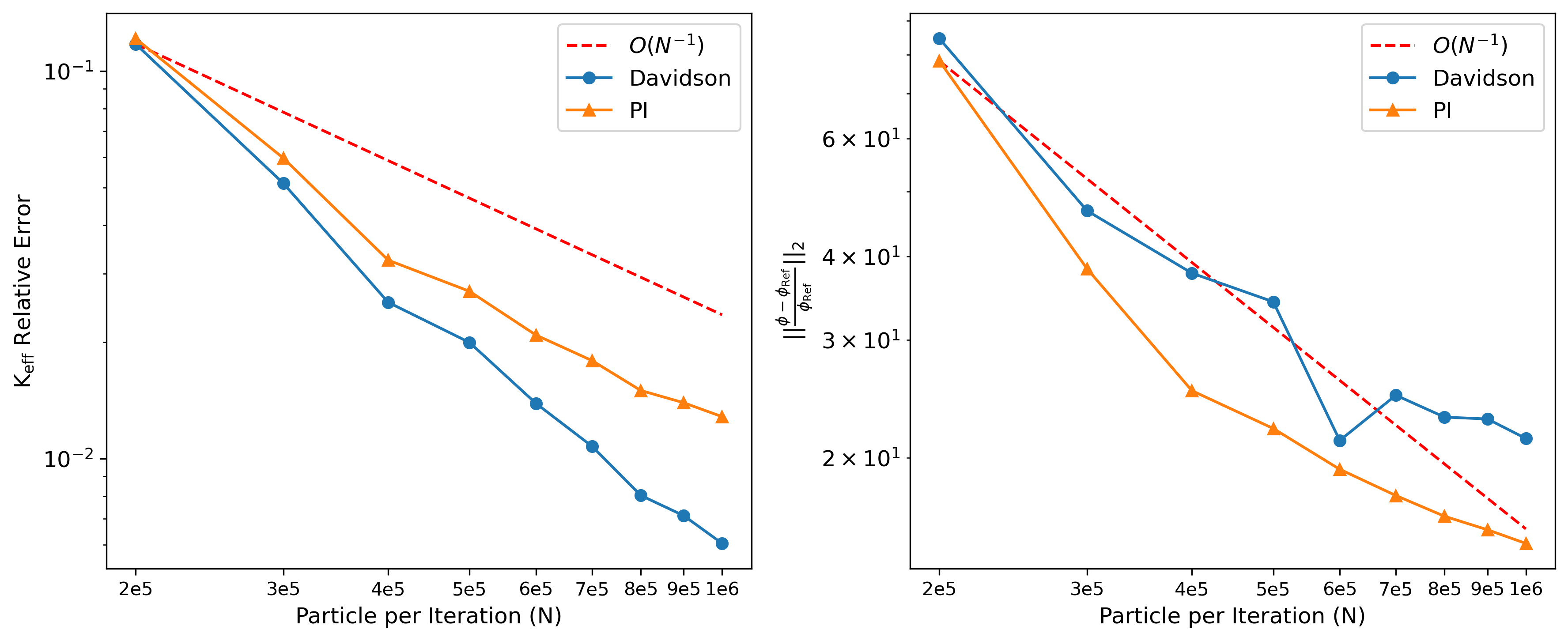}
  \caption{iQMC relative error for $\boldsymbol{k_\mathrm{eff}}$ and scalar flux $\boldsymbol{\phi}$ for varying particle count per iteration $\boldsymbol{N}$, along with the theoretical QMC convergence $\boldsymbol{O(N^{-1})}$.}
  \label{fig:takeda1_error}
\end{figure}

\begin{figure}[!htb]
  \centering
  \includegraphics[scale=0.5]{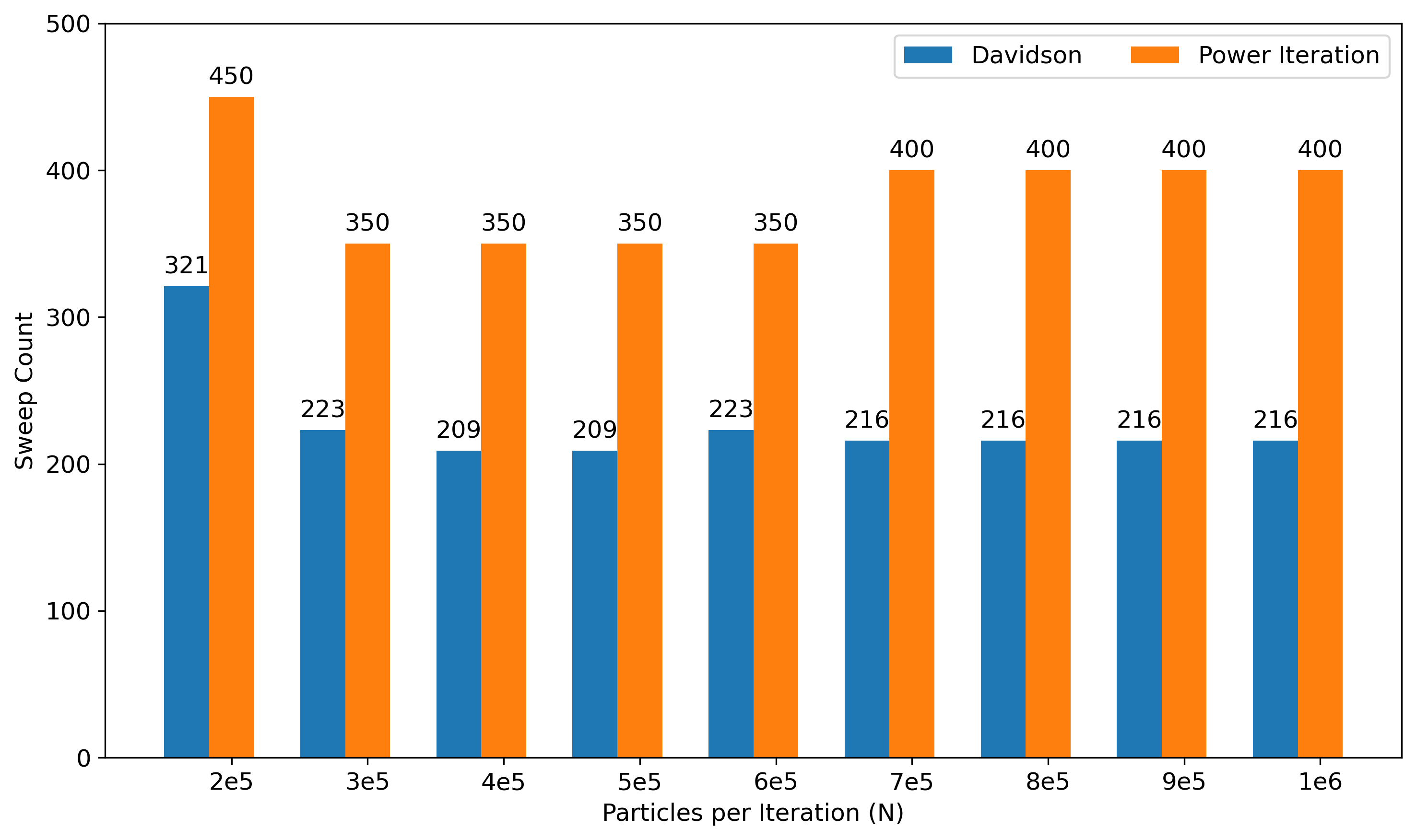}
  \caption{Total number of QMC transport sweeps required for $\boldsymbol{\Delta k_\mathrm{eff}/k_\mathrm{eff} \le 1\times10^{-4}}$ with the Davidson method and power iteration with varying $\boldsymbol{N}$.}
  \label{fig:takeda1_sweeps}
\end{figure}

\begin{figure}[!htb]
  \centering
  \includegraphics[width=\textwidth]{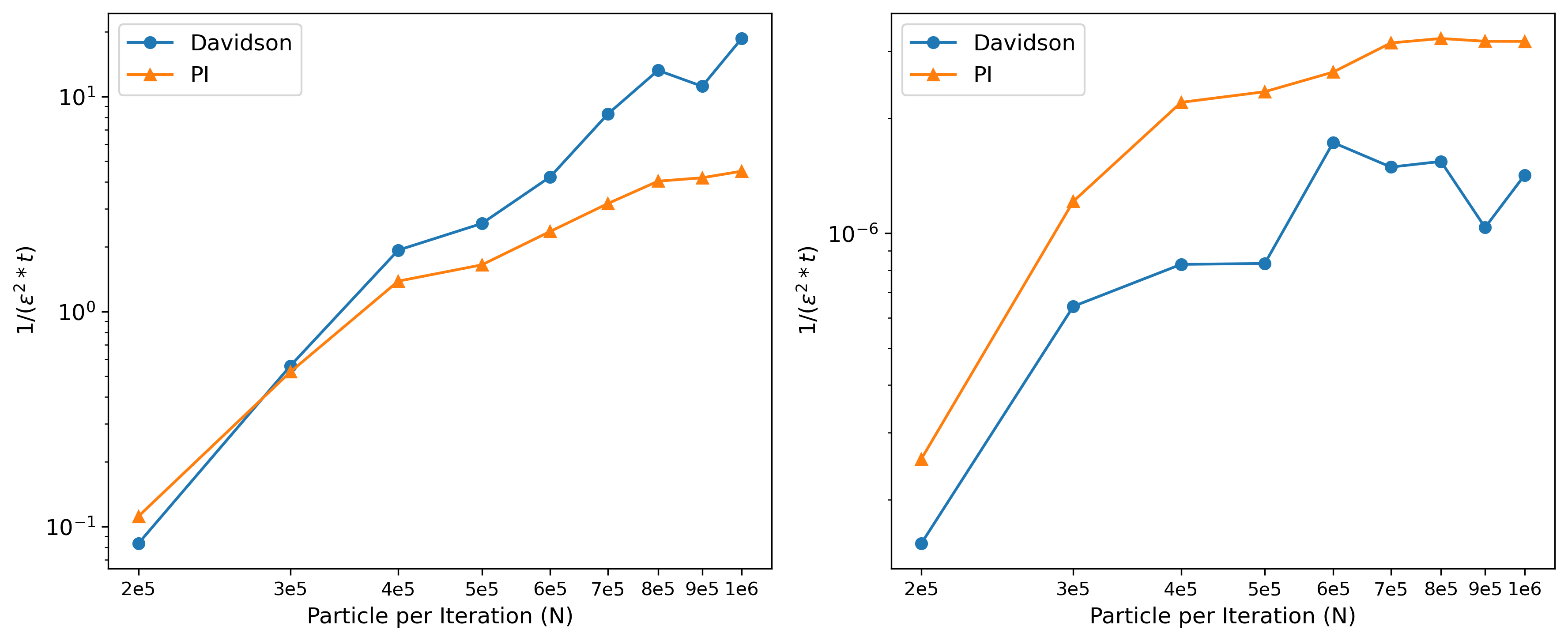}
  \caption{power iteration and Davidson method Figures of Merit (FoM), defined as $\boldsymbol{1/(\epsilon * t^2)}$, where $\boldsymbol{\epsilon}$ is the relative error and $t$ is the simulation time for $\boldsymbol{k_\mathrm{eff}}$ and scalar flux $\boldsymbol{\phi}$ for varying particle count per iteration $\boldsymbol{N}$.}
  \label{fig:takeda1_fom}
\end{figure}

 Figure~\ref{fig:takeda1_error} shows the relative error for both $k_\mathrm{eff}$ and $\phi$ with varying particles per iteration $N$ and the expected $O(N^{-1})$ convergence is achieved for both iterative methods. The scalar flux convergence however, begins to plateau with higher particle count. This is likely due to the spatial error from a relatively coarse mesh and the use of a piece-wise constant source \cite{Pasmann2023}. Figure~\ref{fig:takeda1_error} also shows the Davidson method achieving a more accurate solution for $k_\mathrm{eff}$ but less accurate scalar flux approximation than the power iteration. This is a result of doing a very rough solve of the scattering source using the preconditioner. In contrast, the power iteration solves the scattering source precisely. Figure~\ref{fig:takeda1_sweeps} shows the Davidson method requires nearly $1/2$ the transport sweeps as the power iteration. This directly translates to computational time as the QMC transport sweeps represent the bulk of the computational cost. These results are reflected in Figure~\ref{fig:takeda1_fom} which plots the Figure of Merit, defined as $1/(\epsilon * t^2)$, where $\epsilon$ is the relative error and $t$ is the simulation time. 

\clearpage

\section{CONCLUSIONS}
We demonstrate implementation of the iterative-Quasi-Monte Carlo (iQMC) method to solve k-eigenvalue problems in the Monte Carlo Dynamic Code. Results are given from the 3-D, 2-group, Takeda-1 Benchmark problem for both the power iteration and generalized Davidson method. iQMC achieved the expected $O(N^{-1})$ convergence rate of the scalar flux with both eigenvalue solvers. The Davidson method is shown to be a valuable tool for solving for the dominant eigenvalue, requiring as few as 1/2 the number of transport sweeps for convergence of the eigenvalue than the power iteration. Power iteration performance may be improved by replacing the source iteration with a Krylov method like GMRES or BiCGSTAB. Nonetheless, the power iteration holds a higher figure of merit with regards to the scalar flux thanks to the more precise inner solve. The Davidson method's scalar flux approximation may be improved by increasing the number of preconditioner sweeps, although this would also increase computation time.

\section*{ACKNOWLEDGEMENTS}
This work was funded by the Center for Exascale Monte-
Carlo Neutron Transport (CEMeNT) a PSAAP-III project
funded by the Department of Energy, grant number: DE-
NA003967 and the National Science Foundation, grant number DMS-1906446.

\bibliographystyle{mc2023}
\bibliography{mc2023}

\begin{thebibliography}{10}
\newcommand{\enquote}[1]{``#1''}
\providecommand{\url}[1]{\texttt{#1}}
\providecommand{\urlprefix}{URL }

\bibitem{bickel2009monte}
P.~Bickel, P.~Diggle, S.~Feinberg, U.~Gather, I.~Olkin, and S.~Zeger.
\newblock \emph{Monte Carlo and Quasi-Monte Carlo Sampling}.
\newblock Springer (2009).
\newblock \urlprefix\url{http://www.springer.com/series/692}.

\bibitem{spanier1995quasi}
J.~Spanier.
\newblock \enquote{Quasi-Monte Carlo Methods for Particle Transport Problems.}
\newblock \emph{Monte Carlo and Quasi-Monte Carlo Methods in Scientific
  Computing}, pp. 121--148 (1995).

\bibitem{Pasmann2023}
S.~Pasmann, I.~Variansyah, C.~T. Kelley, and R.~McClarren.
\newblock \enquote{A Quasi–Monte Carlo Method With Krylov Linear Solvers for
  Multigroup Neutron Transport Simulations.}
\newblock \emph{https://doiorg/101080/0029563920222143704}, pp. 1--15 (2023).
\newblock
  \urlprefix\url{https://www.tandfonline.com/doi/abs/10.1080/00295639.2022.2143704}.

\bibitem{Subramanian2011}
C.~Subramanian, S.~V. Criekingen, V.~Heuveline, F.~Nataf, and P.~Havé.
\newblock \enquote{The Davidson Method as an Alternative to Power Iterations
  for Criticality Calculations.}
\newblock American Nuclear Society (2011).

\bibitem{Takeda1991}
T.~Takeda and H.~Ikeda.
\newblock \enquote{3-D Neutron Transport Benchmarks.}
\newblock \emph{Journal of Nuclear Science and Technology}, \textbf{volume~28},
  pp. 656--669 (1991).

\bibitem{variansyah_mc23_mcdc}
I.~Variansyah, J.~P. Morgan, J.~Northrop, K.~E. Niemeyer, and R.~G. McClarren.
\newblock \enquote{Development of MC/DC: a performant, scalable, and portable
  Python-based Monte Carlo neutron transport code.}
\newblock In \emph{International Conference on Mathematics and Computational
  Methods Applied to Nuclear Science and Engineering}. Niagara Falls, Ontario,
  Canada (2023).

\bibitem{Lewis1984}
E.~Miller and W.~J. Lewis.
\newblock \emph{Computational Methods of Neutron Transport}.
\newblock John Wiley and Sons (1984).

\bibitem{hamilton2011numerical}
S.~P. Hamilton.
\newblock \emph{Numerical Solution of the k-Eigenvalue Problem}.
\newblock Ph.D. thesis, Emory University (2011).

\bibitem{warsa2004krylov}
J.~S. Warsa, T.~A. Wareing, J.~E. Morel, J.~M. McGhee, and R.~B. Lehoucq.
\newblock \enquote{Krylov Subspace Iterations for Deterministic k-Eigenvalue
  Calculations.}
\newblock \emph{Nuclear Science and Engineering}, \textbf{volume 147}(1), pp.
  26--42 (2004).

\bibitem{Davidson1975}
E.~R. Davidson.
\newblock \enquote{The Iterative Calculation of a Few of the Lowest Eigenvalues
  and Corresponding Eigenvectors of Large Real-Symmetric Matrices.}
\newblock \emph{Journal of Computational Physics}, \textbf{volume~17}, pp.
  87--94 (1975).

\bibitem{kelley1995iterative}
C.~T. Kelley.
\newblock \emph{Iterative Methods for Linear and Nonlinear Equations}.
\newblock SIAM (1995).

\bibitem{lam2015numba}
S.~K. Lam, A.~Pitrou, and S.~Seibert.
\newblock \enquote{Numba: A llvm-based python jit compiler.}
\newblock In \emph{Proceedings of the Second Workshop on the LLVM Compiler
  Infrastructure in HPC}, pp. 1--6 (2015).

\bibitem{Criekingen2007}
S.~V. Criekingen.
\newblock \enquote{A 2-D/3-D Cartesian Geometry Non-Conforming Spherical
  Harmonic Neutron Transport Solver.}
\newblock \emph{Annals of Nuclear Energy}, \textbf{volume~34}, pp. 177--187
  (2007).
\newblock \urlprefix\url{www.elsevier.com/locate/anucene}.

\end{thebibliography}

\end{document}